# Molecular traffic control in single-file networks with fast catalysts

A. Brzank[1], G. M. Schütz[2], P. Bräuer[1], J. Kärger[1]

[1]*Fakultät für Physik und Geowissenschaft, Universität Leipzig, Abt. Grenzflächenphysik, Linnéstrasse 5, D-04103 Leipzig Germany*

[2]*Institut für Festkörperforschung, Forschungszentrum Jülich, D-52425 Jülich, Germany*

As a model for molecular traffic control (MTC) we investigate the diffusion of hard core particles in crossed single-file systems. We consider a square lattice of single-files being connected to external reservoirs. The (vertical) α-channels, carrying only A-particles, are connected to reservoirs with constant density $\rho_A$. B-particles move along the (horizontal) β-channels, which are connected to reservoirs of density $\rho_B$. We allow the irreversible transition A→B at intersections. We are interested in the stationary density profile in the α- and β-channels, which is the distribution of the occupation probabilities over the lattice. We calculate the stationary currents of the system and show that for sufficiently long channels the currents (as a function of the reservoir densities) show in the limit of large transition rates non analytic behavior. The results obtained by direct solution of the master equation are verified by kinetic Monte Carlo simulations.

## 1. Introduction

The concept of molecular traffic control was introduced in the beginning of the eighties [1, 2]. It is postulated that the effective reactivity of microporous catalysts may be enhanced by directing the reactant and product molecules along different pathways. The so called molecular traffic control effect has recently been verified by Monte Carlo simulations [3]. For a better insight into the restrictions and limitations of the MTC effect it is desirable to understand the conditions which determine the actual flow of particles. This question is closely related to the problem of calculating the stationary density profile of the system. From a theoretical point of view the MTC system can be modelled as an array of several interacting symmetric exclusion processes (SEP) [4]. The stationary properties of the SEP [5] within a channel, in particular, the linear density profile with the stationary particle current $j \propto 1/L$ together with the transition rules at the catalytic sites leed to an interesting and unexpected behavior of the composed system.

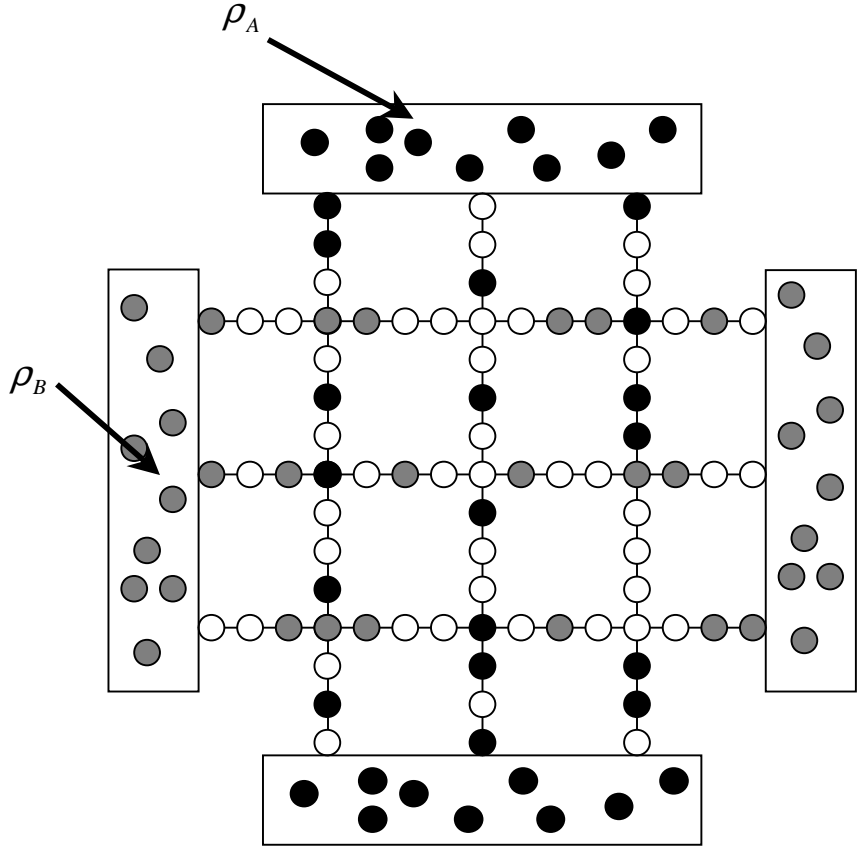

**Fig. 1** MTC system with three α-channels (vertical) and three β-channels (horizontal) and $L=3$ lattice sites between adjacent intersections.

## 2. Description of the system

We consider $2N$-1 channels of type α and β. Slightly generalizing the set-up of [6] we assume the α(β)-channels to be connected to reservoirs of constant densities such that the entrances of the respective channels have fixed particle densities $\rho_{A(B)}$. There should be no interaction other than hard core repulsion which forbids double-occupancy of the lattice sites. $D$ is the elementary (attempt) rate of hopping between lattice sites and it is assumed to be the same for both species of particles. Therefore we have a network of interacting one-dimensional symmetric exclusion processes (SEP). From this we can expect linear density profiles between two intersections [4], the slope being inversely proportional to the number of lattice sites $L$. $L$ is the number of lattice sites between two neighboring intersections or intersection and the adjacent boundary site.

At the intersections we place a catalyst that changes irreversibly A-particles into B-particles. In a discrete-time setting this mechanism can be specified by saying that per time unit $\Delta t$ an A-particle is converted into a B-particle without having a jump attempt. In a continuous-time description this probability reduces to the reaction rate $c$ which is the reaction probability per time unit $\Delta t$.

In our present calculations we assume $L \gg 1$ and $c$ large enough so it is unlikely that an A-particle that has entered an intersection point will leave the intersection before having been converted into a B-particle. Therefore in the steady state the bulk of the network contains almost no A-particles. Only the first and the last β-channels and the first segment of the α-channels are of interest. Due to reflection symmetry it is sufficient to consider only a half-system and finally the system reduces to Fig. 2.

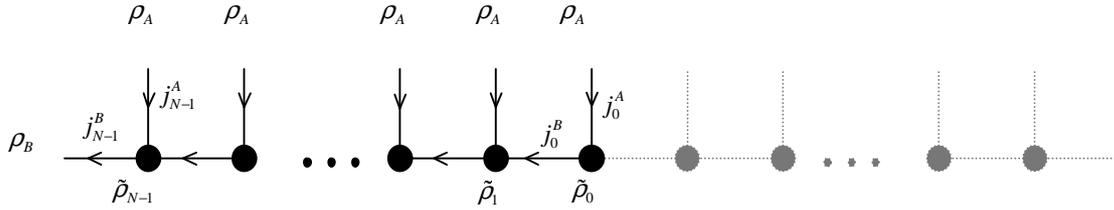

**Fig. 2** Reduced MTC system. The B-particle density of the $r$'th intersection is labled with $\tilde{\rho}_r$.

## 3. The general case

Let us label the lattice sites for the $r$'th α-channel segment from 1 to $L$, denote the catalytic site by $I=L+1$ and the boundary site by 0. We use a similar labeling for the $r$'th β-channel (bulksites 1 to $L$, intersections by $I=0$) since it will be clear which one we refer to. Because of the exclusion interaction the local occupation numbers $a_k^r$ and $b_k^r$ for A- and B-particles can only take values 0,1. We also define the "vacancy occupation number" $v_k^r$. In the α-channel we have $v_k^r = 1 - a_k^r$, in the β-channel $v_k^r = 1 - b_k^r$ and on the catalytic site $v_I^r = 1 - b_I^r - a_I^r$. The expected local densities $\langle a_i^r \rangle$ for A-particles and $\langle b_i^r \rangle$ for B-particles satisfy the following set of equations

$$\frac{d}{dt}\langle a_1^r \rangle = D\left(\langle a_2^r \rangle - \langle a_1^r \rangle + \rho_A\left(1 - \langle a_1^r \rangle\right) - \left(1 - \rho_A\right)\langle a_1^r \rangle\right) \tag{3.1}$$

$$\frac{d}{dt}\langle a_i^r \rangle = D\left(\langle a_{i+1}^r \rangle + \langle a_{i-1}^r \rangle - 2\langle a_i^r \rangle\right) \qquad 2 \leq i \leq L-1 \tag{3.2}$$

$$\frac{d}{dt}\langle a_L^r \rangle = D\left(\langle a_{L-1}^r \rangle - \langle a_L^r \rangle + \langle a_I^r v_L^r \rangle - \langle a_L^r v_I^r \rangle\right) \tag{3.3}$$

$$\frac{d}{dt}\langle a_I^r \rangle = D\left(\langle a_L^r v_I^r \rangle - \langle a_I^r v_L^r \rangle\right) - c\langle a_I^r \rangle \tag{3.4}$$

$$\frac{d}{dt}\langle b_I^r \rangle \equiv D\left(\langle b_L^{r-1} v_I^r \rangle - \langle b_I^r v_L^{r-1} \rangle + \langle b_1^r v_I^r \rangle - \langle b_I^r v_1^r \rangle\right) + c\langle a_I^r \rangle \tag{3.5}$$

$$\frac{d}{dt}\langle b_1^r \rangle = D\left(\langle b_2^r \rangle - \langle b_1^r \rangle + \langle b_I^r v_1^r \rangle - \langle b_1^r v_I^r \rangle\right) \tag{3.6}$$

$$\frac{d}{dt}\langle b_i^r \rangle = D\left(\langle b_{i+1}^r \rangle + \langle b_{i-1}^r \rangle - 2\langle b_i^r \rangle\right) \quad 2 \leq i \leq L-1 \tag{3.7}$$

$$\frac{d}{dt}\langle b_L^r \rangle = D\left(\langle b_{L-1}^r \rangle - \langle b_L^r \rangle + \langle b_I^{r+1} v_L^r \rangle - \langle b_L^r v_I^{r+1} \rangle\right) \tag{3.8}$$

$$\frac{d}{dt}\langle b_L^N \rangle = D\left(\langle b_{L-1}^N \rangle - \langle b_L^N \rangle\right) + \rho_B\left(1 - \langle b_L^N \rangle\right) - (1-\rho_B)\langle b_L^N \rangle. \tag{3.9}$$

In these equations $0 \leq r \leq N-1$, notice that $\langle b_{L+1}^N \rangle \equiv \rho_B$ and $\langle a_0^r \rangle \equiv \rho_A$. The linear nature of the bulk equations (3.2), (3.7) results from a cancellation of joint probabilities $\langle a_i^r a_{i+1}^r \rangle$, $\langle b_i^r b_{i+1}^r \rangle$ due to the SU(2) symmetry of the symmetric exclusion process [7]. For a generic channel they may be written in the standard form of a lattice continuity equation for the local density

$$\frac{d}{dt}\rho_n = j_{n-1} - j_n \tag{3.10}$$

with the densities $\rho_n^A = \langle a_n \rangle$ or $\rho_n^B = \langle b_n \rangle$ respectively at site n and the corresponding bulk currents

$$j_n^A = D\left(\langle a_{n-1} \rangle - \langle a_n \rangle\right) \tag{3.11}$$

$$j_n^B = D\left(\langle b_{n-1} \rangle - \langle b_n \rangle\right) \tag{3.12}$$

between bonds (*n*, *n*+1).

For the stationary solution that we wish to study the time derivative vanishes. Because of particle conservation in the bulk the currents $j_n^A$, $j_n^B$ do not depend on the lattice site *n* which implies linear density profiles. For notational simplicity we will drop the *n* and add the information about the channel we are talking about instead.

For $c \neq 0$ we derive from (3.3)-(3.6) for the r'th channel

$$j_r^A = c\langle a_I^r \rangle \tag{3.13}$$

$$j_r^B = j_{r-1}^B + c\langle a_I^r \rangle; \qquad j_0^B = -j_0^B + c\langle a_I^0 \rangle \tag{3.14}$$

(In the absence of catalysis ($c=0$) one would have an equilibrium state with all currents equal to zero). (3.13), (3.14) imply conservation of currents

$$j_r^B = j_{r-1}^B + j_r^A \qquad 1 \leq r \leq N-1 \tag{3.15}$$

$$j_0^B = \tfrac{1}{2} j_0^A. \tag{3.16}$$

For futher analysis we neglect correlations between the occupancy of a catalytic site and its three neighboring sites. This mean field approximation is motivated by exact results for the correlations in the stationary state of the symmetric exclusion process from which it is known [8] that nearest neighbour correlations in the vicinity of the boundary of a system of size $L$ are of order $1/L^2$ and hence small under the assumption $L \gg 1$ made here. Within mean field we therefore replace joint probabilities $\langle xy \rangle$ by the product $\langle x \rangle \langle y \rangle$. Eq. (3.11) together with (3.3) and Eq. (3.12) together with (3.6) then become two equations for the currents depending only on the local densities of the bordering catalytic sites $\tilde{\rho}_r^A$, $\tilde{\rho}_r^B$, $\tilde{\rho}_{r+1}^B$.

$$j_r^A = D \frac{\tilde{\rho}_r^A + \rho_A(\tilde{\rho}_r^B - 1)}{L(\tilde{\rho}_r^B - 1) - 1} \qquad 0 \leq r \leq N-1 \tag{3.16}$$

$$j_r^B = D \frac{\tilde{\rho}_r^B + \tilde{\rho}_{r+1}^B(\tilde{\rho}_r^A - 1)}{L(1 - \tilde{\rho}_r^A) + 1} \qquad 0 \leq r \leq N-1. \tag{3.17}$$

For further discussion we neglegt the densities of A-particles on the intersections which comes from the model assumption that an A-particle is very likely to be converted into a B-particle before returning to the α-channel. (3.16) and (3.17), together with (3.15), finally leads to a set of equations for the unknown densities $\tilde{\rho}_r^B \equiv \tilde{\rho}_r$.

$$\frac{\tilde{\rho}_r - \tilde{\rho}_{r+1}}{L+1} = \frac{\tilde{\rho}_{r-1} - \tilde{\rho}_r}{L+1} + \frac{\rho_A(\tilde{\rho}_r - 1)}{L(\tilde{\rho}_r - 1) - 1} \qquad 0 \leq r \leq N-1 \tag{3.18}$$

$$\frac{\tilde{\rho}_0 - \tilde{\rho}_1}{L+1} = \frac{1}{2} \frac{\rho_A(\tilde{\rho}_0 - 1)}{L(\tilde{\rho}_0 - 1) - 1} \tag{3.19}$$

These equations are quadratic in $\tilde{\rho}_r$ and to leading order in the system size one has the solutions

$$\tilde{\rho}_r = \begin{cases} 1 \\ \tfrac{1}{2}(\tilde{\rho}_{r+1} + \tilde{\rho}_{r-1} + \rho_A); \quad \tilde{\rho}_0 = \tilde{\rho}_1 + \tfrac{1}{2}\rho_A \end{cases} \qquad 1 \leq r \leq N-1 \tag{3.20}$$

Before giving an explicit expression for the local densities at the intersections we would like to discuss which of these solutions the system selects. We note that due to exclusion $\tilde{\rho}_r \leq 1$. Due to particle conservation in the bulk of the system all currents are non-negative. This implies an increase of the densities $\tilde{\rho}_r$ for decreasing $r$. Therefore all solutions with $\tilde{\rho}_n = 1$ and $\tilde{\rho}_m < 1$ for $m<n$ need to be discarded. The actually selected solution is determined by the boundary densities $\rho_A$ and $\rho_B$. For small $\rho_A$ each α-channel is in its maximal-current state since the probability of finding A-particles on the catalytic sites is almost zero. Each β-channel can support the accumulated currents that enter throug the α-channels. However, this implies a rather high density of B-particles at the first catalytic site. Then, with an adiabatic increase of $\rho_A$ there is a critical value at which this density reaches 1. Increasing $\rho_A$ further eventually leads to the second catalytic site saturating. Then there is no current any more in the first part of the β-channel. This goes on until also the last catalytic site saturates and the system reaches a situation where the current is limited by the capacity of the last segment of the β-channel and only in this channel there is an actual flow of particles.

To quantify this we note again that for small $\rho_A$ none of the catalytic sites is saturated and the recursion relation (3.20) is solved by

$$\tilde{\rho}_r = \tfrac{1}{2}\left(N^2 - r^2\right)\rho_A + \rho_B. \tag{3.21}$$

Notice that $\tilde{\rho}_r \leq 1$ which implies the range

$$0 \leq \rho_A < 2\frac{1-\rho_B}{N^2}. \tag{3.22}$$

Increasing $\rho_A$ to such an extent that the densities of the first $s$ intersections become 1 we have

$$\tilde{\rho}_r = \frac{1}{N-s}\left[(N-r)+(r-s)\rho_B + \frac{r-s}{2}(N-r)(N-s)\rho_A\right] \quad r = s+1..N-1 \tag{3.23}$$

with the corresponding range

$$\rho_A < \frac{2(1-\rho_B)}{(N-s-1)(N-s)}. \tag{3.24}$$

## 4. Simulations and Results

We used dynamic Monte Carlo simulations to verify the results obtained in the previous section. The algorithm uses random sequential update. Statistical errors are of order 0.1% and therefore not additionally marked on the graphs. We calculated the stationary profile for a system of $N=1$ $\alpha$-channels and $L=250$ and a system of $N=3$ $\alpha$-channels and $L=250$. In the first case we find the two solutions

$$j = \begin{cases} \dfrac{D}{2}\dfrac{\rho_A}{L} & 0 \leq \rho_A \leq 2-2\rho_B \\ D\dfrac{1-\rho_B}{L} & 2-2\rho_B < \rho_A \leq 1 \end{cases} \qquad (4.24)$$

which implies that the system selects the smaller of the two values (4.24).

$$j = \min\left\{\dfrac{D}{2}\dfrac{\rho_A}{L} \equiv j_{\max}^A, D\dfrac{1-\rho_B}{L} \equiv j_{\max}^B\right\} \qquad (4.25)$$

This has an intuitive physical meaning. In each channel the system tries to maximize its current, but the actually selected current is limited by the smaller one of the two maximal currents. To understand the origin of this selection principle we employ the picture developed in the preceding section. Notice that the stationary probability of finding an A-particle on the catalytic site is almost zero. Now let us assume that first $\rho_A$, i.e. the reservoir density of A-particles, is very small. For small $\rho_A$ the system tries to sustain the current $j = j_{\max}^A$ which is possible as long as the $\beta$-channel can support this current. This is the case if the density of B-particles at site $N$ required to generate this current in the $\beta$-channel is less than 1. Notice that in this regime the density of A-particles both on site $I$ and on the last site of the $\alpha$-channel is nearly zero. Now we assume $\rho_A$ to be increased adiabatically, i.e., so slowly that the system reaches stationarity before a further increase occurs. On the catalytic site the B-density $\tilde{\rho}$ increases in order to sustain the enhanced stationary current. If $\rho_A$ becomes so large that $j_{\max}^A = j_{\max}^B$ one has $\tilde{\rho} = 1$. Then the current is limited by the capacity of the $\beta$-channel and increasing $\rho_A$ further does not increase the current. Instead the last site $L$ in the $\alpha$-channel acquires a finite density of A-particles.

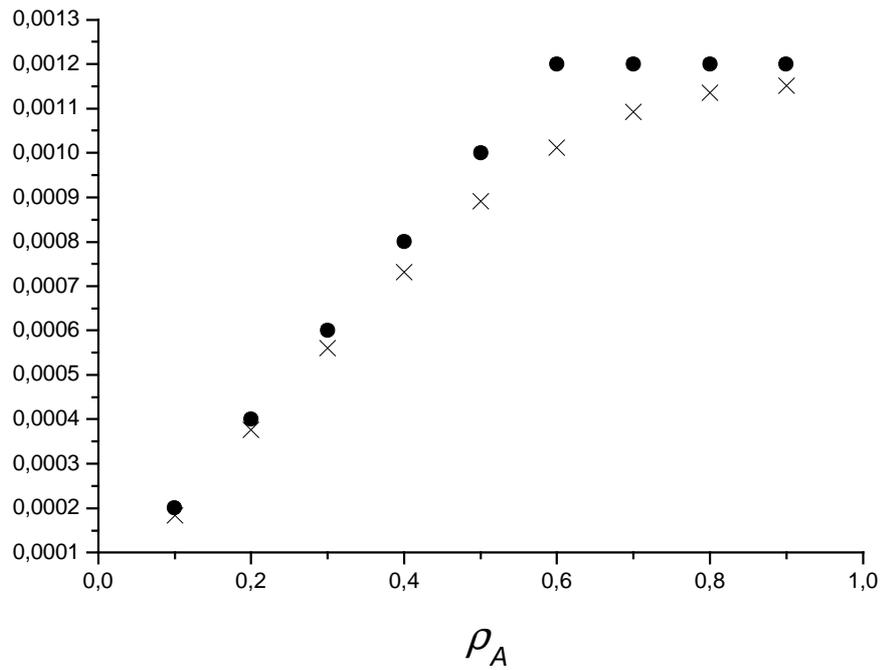

**Fig. 3** Particle current obtained by Monte carlo simulation (x) vs. theoretical current (dots). Lattice $N=1$, $L=250$, $\rho_B = 0.7$. The deviations are mainly due to finite-size corrections.

For increasing $L$ one expects the simulations to approach the theoretical densities. Fig. 4 shows the calculations for a lattice of $N=1$, $\rho_A = 0.6$ and $\rho_B = 0.7$ with the theoretical density at the intersection $\tilde{\rho} = 1$.

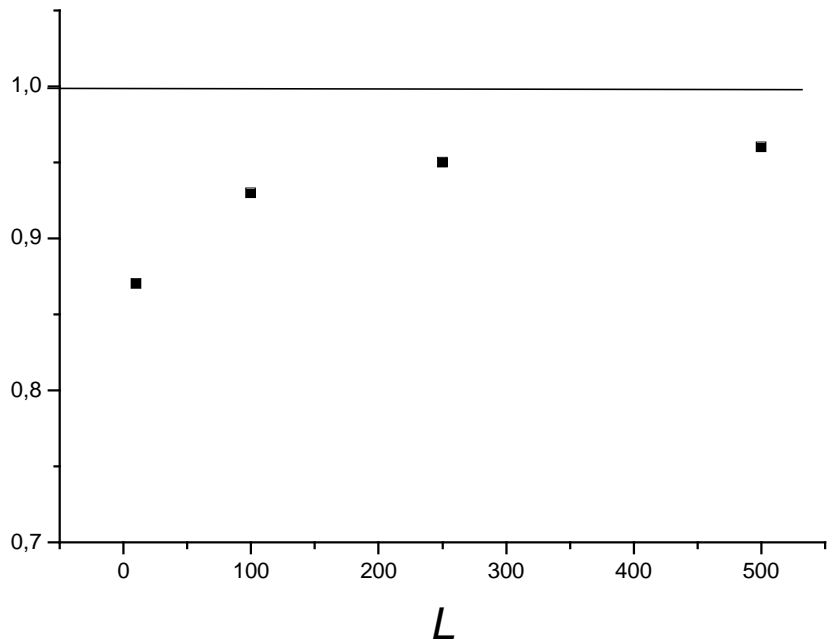

**Fig. 4** For a lattice $N=1$, $\rho_A = 0.6$, $\rho_B = 0.7$ one findes $\tilde{\rho} = 1$. For increasing $L$ the density of the intersection obtained by simulations approaches the theoretical value.

We consider the system of $N=3$ α-channels. For simplification we assume $\rho_B$ to be zero. Fig. 5 shows the theoretical currents of the α-channels and β-channels as a function of $\rho_A$. The first intersection $\tilde{\rho}_0$ saturates when $\rho_A = \frac{2}{9}$, $\tilde{\rho}_1$ becomes one for $\rho_A = \frac{1}{3}$. For the same lattice Fig. 6 compares the densities obtained by simulation with the theoretical values.

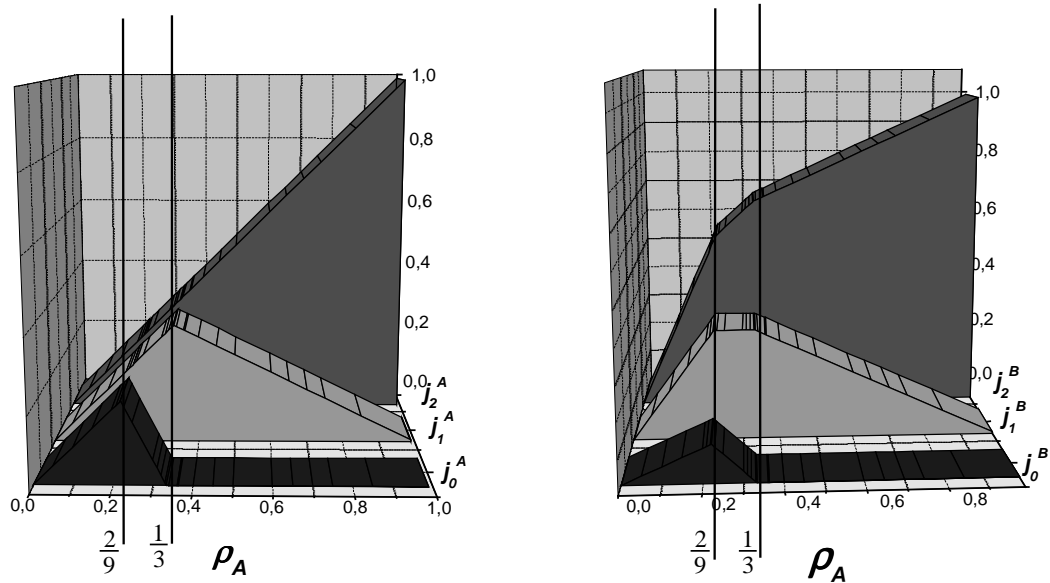

**Fig. 5** Normalized theoretical currents of the α-channels (left), β-channels (right) in a lattice of $N=3$. The first intersection $\tilde{\rho}_0$ saturates when $\rho_A = \frac{2}{9}$, $\tilde{\rho}_1$ becomes one for $\rho_A = \frac{1}{3}$.

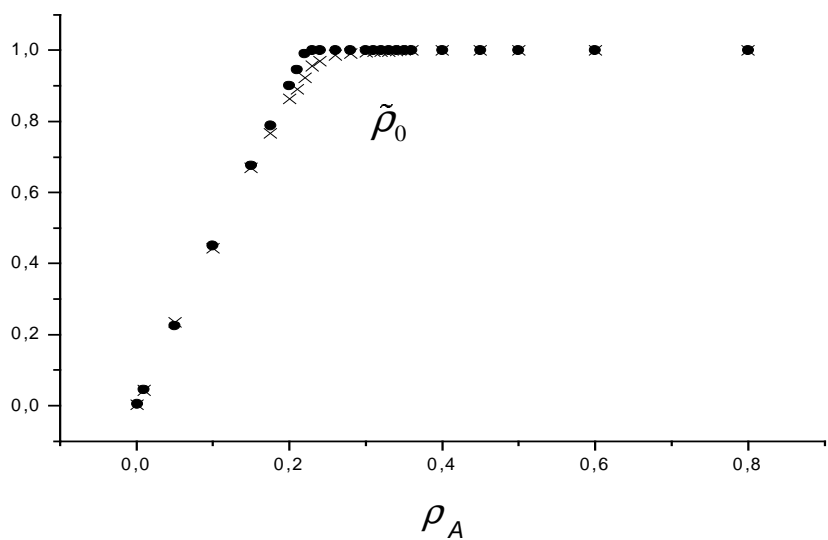

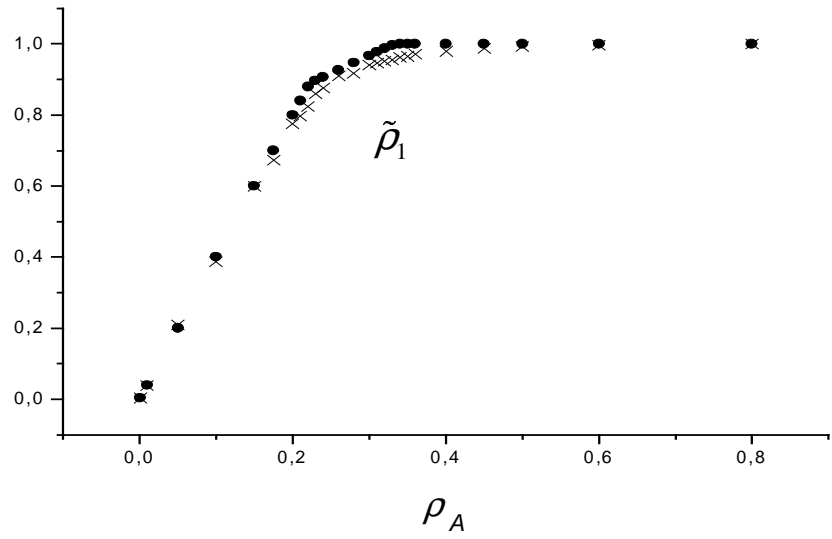

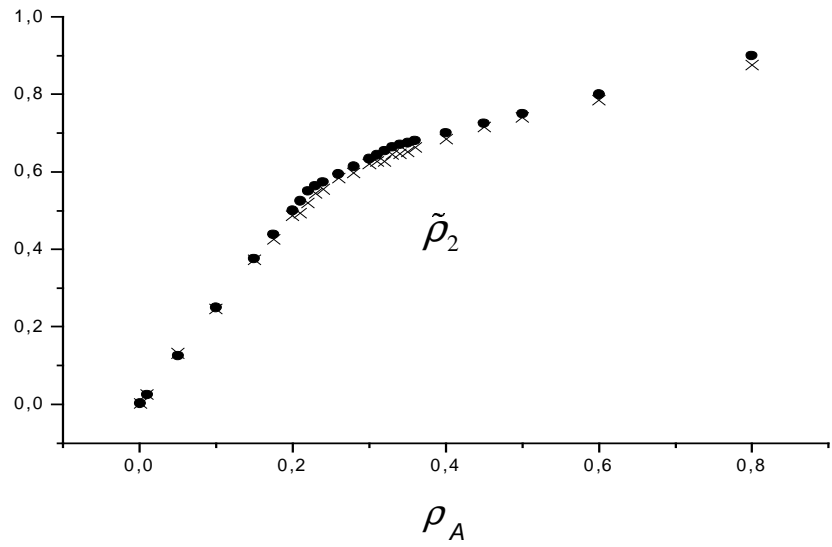

**Fig. 6** Densities $\tilde{\rho}_0$, $\tilde{\rho}_1$, $\tilde{\rho}_2$ of B-particles obtained by Monte Carlo simulation (x) vs. theoretical densities (dots). Lattice $N=3$, $L=250$, $\rho_B = 0$.

## 5. Conclusion

We discussed the stationary occupation probabilities of the molecular traffic control system with fast catalysts. Our main results are the obtained intersection densities (3.21) - (3.24), calculated in the limit of large file-length. They are sufficient for the derivation of the density profile as a function of the reservoir densities $\rho_A, \rho_B$ due to linear density profiles of the channel segments. As an intriguing result we showed that an increase of the reservoir gradient not necessarily means an increase of the currents inside the system. In fact due to saturating intersection sites certain channels show (Fig. 4) an actual decrease of the current and finally become zero. Since the mean field approximation is essential in our calculation the results are not valid for systems of short file length. The stationary behavior of these systems remains an open question. The main ingredient in our approach is current conservation inside channels. Hence a similar analysis may be performed for more realistic model systems with density-dependent currents.